\documentclass[aps,prl,twocolumn,superscriptaddress,showpacs]{revtex4}

\usepackage{amsmath}
\usepackage{graphicx}

\begin{document}
\title{Edge excitations and Topological orders in
rotating Bose gases}
\author{M.~A. Cazalilla}
\affiliation{Donostia International Physics Center (DIPC),
Manuel de Lardizabal 4, 20018-Donostia, Spain.}
\author{N. Barber\'an}
\affiliation{Dept.  ECM, Facultat de F\' isica, 
Universitat de Barcelona. Diagonal 647, 08028-Barcelona, Spain.}
\author{N.~R. Cooper}
\affiliation{Cavendish Laboratory, Madingley Road, Cambridge, CB3 0HE, United Kingdom.}
\begin{abstract}
The edge excitations and related topological orders of correlated states
of a fast rotating Bose gas are  studied. Using exact diagonalization of small systems, 
we compute the energies and number of edge excitations, as well as the 
boson occupancy near the edge for various states. 
The chiral Luttinger-liquid theory of Wen is found to be a good description
of the edges of the bosonic Laughlin and other states identified as members
of the principal Jain sequence for bosons.   However,  we  find that in a harmonic trap
the edge of the state identified as the Moore-Read (Pfaffian) state shows a number of anomalies.  An experimental way of detecting these correlated states
is also discussed.
\end{abstract}
\maketitle

It has been  argued~\cite{WG00,CWG01,B04} that quantum fluctuations
can destroy a Bose-Einstein condensate if it rotates very fast. 
Since large amounts of angular momentum  can
be imparted to a cold atomic gas, experimentalists have been able
to create  systems with a  large  
number of  vortices, $N_v$~\cite{SC04,Bt04,B04}. 
Thus, the question of what happens when $N_v$  
eventually becomes comparable to the number  of particles, 
$N$, has been raised~\cite{WG00,CWG01}.
In ~\cite{CWG01}, it was shown that  
for $\nu = N/N_v \sim 10$, the BEC and the Abrikosov
vortex lattice  are destroyed and replaced by
a series of ``vortex liquid'' states, some of  which 
are  \emph{incompressible} and
exhibit  good overlap with bosonic versions of  
wavefunctions known from fractional quantum Hall 
effect (FQHE). 

 Good overlap is usually a strong indication that 
some of the correlations  of a state obtained from 
exact diagonalization are captured by a model wavefunction. However, 
Wen~\cite{W95} has  emphasized that FQHE states possess 
a new type of \emph{quantum} order, which he dubbed 
`topological order', that provides a better way to classify them. In a deep sense, 
topological order can be regarded as a measure of 
the quantum entanglement existing 
between the particles in a correlated quantum Hall (QH) state~\cite{W02}.

 In this paper we study the topological order of the vortex liquids
as reflected in their edge properties~\cite{W95}.  
\emph{In the rotating frame}, edge excitations
are the low-lying excitations of the vortex liquid~\cite{C03} and, contrary to the 
ground  states~\cite{WC99,WG00,CWG01,RJ03}, so far they  have
received little attention.  Based on the  strong similarities with 
electron FQHE physics, chiral Luttinger liquids~\cite{W92} and 
similar edge excitations~\cite{W95} are expected, but an explicit 
demonstration is lacking for a harmonically confined gas of bosons under 
rotation. This is provided here by numerically diagonalizing  the Hamiltonian
of small systems.

The topological order can be studied 
at the edge of  fairly small droplets of QH
liquid~\cite{W92,W95,ZPM96}.  In what follows, we shall focus on three  types 
of states that have been identified in previous works on small droplets~\cite{WC99,WG00}.  
In the high angular momentum end, we study  the edge properties 
of the Laughlin state (corresponding to a bulk filling fraction $\nu = \frac{1}{2}$).  
By decreasing the  total angular momentum,  we come across the
compact composite-fermion states discussed in Ref.~\cite{WC99}. 
Amongst these states, we find evidence for the topological 
order (and related edge structures) 
that correspond to bulk states with filling fractions $\nu  = \frac{2}{3}$ 
and $\nu = \frac{3}{4}$ of the principal Jain sequence.
At even lower angular momentum, we study the edge properties 
of the state identified\cite{WG00} as the 
finite-sized Moore-Read (or Pfaffian) state\cite{MR91} 
($\nu=1$).
We observe a number of anomalies that persist up 
to the largest sizes studied here ($N = 13$).
The bulk properties of the states at these filling fractions have also
been studied in exact diagonalizations in edgeless
geometries~\cite{CWG01,RJ03} which have established their
interpretation in terms of the Laughlin ($\nu=1/2$), composite fermion
($\nu=2/3,3/4$) and Moore-Read ($\nu=1$) states. For the latter,
calculations on the sphere~\cite{RJ03} and on the torus~\cite{CWG01}  
found the correct shift and  ground state degeneracy 
(which are also consequence of the topological order~\cite{W95}).
 
 An ultracold gas that rotates rapidly  in a cylindrically symmetric harmonic trap 
acquires a pancake shape and eventually becomes quasi
two-dimensional (2D) when the chemical potential $\mu < \hbar \omega_{||}$,
where $\omega_{||}$ is the axial trapping frequency~\cite{Ho01}.
Furthermore, the Coriolis force acts as an effective Lorentz
force, which in a quasi-2D system leads to  
Landau levels (LL's) separated by  an energy 
$ 2 \hbar \omega_{\perp}$~\cite{Ho01}. For
$\mu <  2\hbar \omega_{\perp}$, all atoms
lie in the lowest Landau level  (LLL), and the total single-particle energy is proportional (up to a constant) to the angular momentum. 
Here we shall be interested in this limit, which has been already achieved in the
experiments~\cite{SC04}.  In the rotating frame, the Hamiltonian (relative to the zero-point energy) is:
\begin{equation}
H_{\rm LLL} = T + U_2 =  \hbar (\omega_{\perp}-\Omega) L + 
g \sum_{i< j} \delta({\bf r}_i - {\bf r}_j), \label{ham}
\end{equation}
$\hbar L$ being the total axial angular momentum and 
$g = \sqrt{8\pi} \hbar \omega_{\perp} (a_s/\ell_{||}) \ell^2$ the effective 
coupling for a gas harmonically confined to two dimensions ($a_s < \ell_{||}$ 
being the scattering length that characterizes the atom-atom interaction 
in 3D, $\ell_{||} = \sqrt{\hbar/M\omega_{||}}$ the axial oscillator length, 
and $\ell = \sqrt{\hbar/M\omega_{\perp}}$,   $M$ being the atom mass).

{\bf Laughlin state:}  For  $L = L_0 =  N(N-1)$,  the ground state of $H_{\rm LLL}$ 
is~\cite{WG00,WC99,C03}:
\begin{equation}
\Phi_0(z_1, \ldots, z_N) = \big\{\prod_{i< j} (z_i - z_j)^2 \big\} \: e^{-\sum_{i=1}^{N} |z_i|^2/2},
\end{equation}
where $z_j \equiv (x_j + i\: y_j)/\ell$. This wavefunction vanishes when any two particles  coincide
and therefore $U_2 |\Phi_0\rangle = 0$. This property is maintained if it is multiplied by
an arbitrary symmetric polynomial of the $z_i$. Edge excitations of the above 
state are generated~\cite{W92,C03} by  elementary symmetric polynomials of the form 
$s_m = \sum_{i_1 < i_2 < \cdots < i_m} z_{i_1} z_{i_2} \cdots z_{i_m}$. Thus, 
$L$ is increased by $m$ units and, according to (\ref{ham}), the excitation energy is 
$\hbar (\omega_{\perp} - \Omega) m$. In the rotating frame, the 
edge excitations are the lowest
energy excitations of a rotating Bose-gas in the Laughlin state~\cite{C03}. Their degeneracy for
$L = L_0 + m$ is given by the number of distinct ways $m$ can be written as a sum  of
smaller non-negative integers (\emph{i.e.} partitions of $m$, $p(m)$). 
For example, for $m = 4$ there
are five degenerate states: $(s_4, s_3s_1, (s_2)^2, s_2 (s_1)^2,
(s_1)^4 ) \times \Phi_0$.  The properties of these wavefunctions are captured by
an effective field theory~\cite{W92,W95,C03}, which treats these excitations as a non-interacting
phonon system with Hamiltonian  $H_{\rm edge} = \hbar (\omega_{\perp} - \Omega) L $, 
where $L = L_0 + \sum_{m > 0} m \: b^{\dag}_m b_m$, and $[b_m, b^{\dag}_n] = \delta_{mn}$,
commuting otherwise. Hence, the number of edge states (NoS) for a given $m$ can
be obtained from a generating  (or partition) function: $
Z(q) = {\rm Tr} \: q^{L-L_0} \, = \prod_{m>0} ( 1 - q^m)^{-1} = \sum_{m>0} p(m) q^m$.
In table~\ref{table1}, we compare the theoretical NoS  ($N \to \infty$)
with the numerical results, finding excellent agreement for $m \le N$. The deviations 
are due to finite-size effects and can be accounted by a `truncated'  function $Z^N(q)$,
where the product is restricted to $0 < m \le N$.
\begin{table}
\caption{Number of edge states (NoS)  vs. excitation angular momentum  $m$.
In the data for the Moore-Read (Pfaffian) state  $\lambda = 0$ for a \emph{pure} 
three-body interaction and $\lambda = 1$ for a \emph{pure} two-body interaction.
\label{table1}}
\begin{ruledtabular}
\begin{tabular}{l|ccccccc}
$m = L - L_0$ &  0 & 1 & 2 & 3 & 4 & 5 & 6 \\ \hline
Laughlin ($N = 5, \, L = 20$) &  1 & 1 & 2 & 3 & 5 & 7 & 10\\
Laughlin ($N = 6, \, L = 30$) &  1 & 1 & 2 & 3 & 5 & 7 & 11\\
Laughlin ($N\to \infty$) & 1 & 1 & 2 & 3 & 5 & 7 & 11\\
$\{4,2\}$ CF ($N =6, \, L = 20$) & 1 & 2 & 5 & 8  \\
$\{5,2\}$ CF ($N = 7,\, L = 30 $) & 1 & 2 &  5 & 9 & 15   \\
Jain $\nu = \frac{2}{3}$ ($N\to \infty$) & 1 & 2 & 5 & 10 & 20 &  36 & 65 \\
Moore-Read ($\lambda = 0$, $ N = 8, \, L = 24$) & 1 & 1 & 3 & 5 & 10 & 15   \\
Moore-Read ($\lambda = 0$, even $N \to \infty$) & 1 & 1 & 3 & 5 & 10 & 16 & 28 \\
Moore-Read ($\lambda = 1$,  $N = 12, \, L = 60$)  &1 & 4 & 10 & 21 \\ 
Moore-Read ($\lambda = 0$, $N = 7, \, L = 18$) & 1 & 2 & 4 & 7 & 12 \\
Moore-Read ($\lambda = 0$, odd $N \to \infty$) & 1 & 2 & 4 & 7 & 13 & 21 & 35 \\
Moore-Read ($\lambda = 1$,  $N = 13, \, L = 72$)  &1 & 6 & 14 & $\geq 20$ \\
\end{tabular}
\end{ruledtabular}
\end{table}

Another  prediction of the theory~\cite{C03,W92} is the form of the groundstate boson occupancy,  $n(l)$, 
just below the highest occupied orbital,  $l_{\rm max} = 2(N-1)$. The following
ratios of $n(l)/n(l_{\rm max})$ are predicted: $1:2:3:4$ for $l = l_{\rm max}, \ldots, l_{\rm max} - 3$.
For $N = 7$ bosons in the Laughlin state, we find $1.0: 2.0 : 2.9 : 3.5$, which is in good
agreement with the theory given that its  validity is  limited to 
$l_{\rm max} - l \lesssim \sqrt{N} \sim 3$~\cite{W92}. Similar agreement was found for 
the fermion Laughlin state in Ref.~\cite{W92}.
\begin{figure}[t]
\begin{center}
\includegraphics[width=\columnwidth]{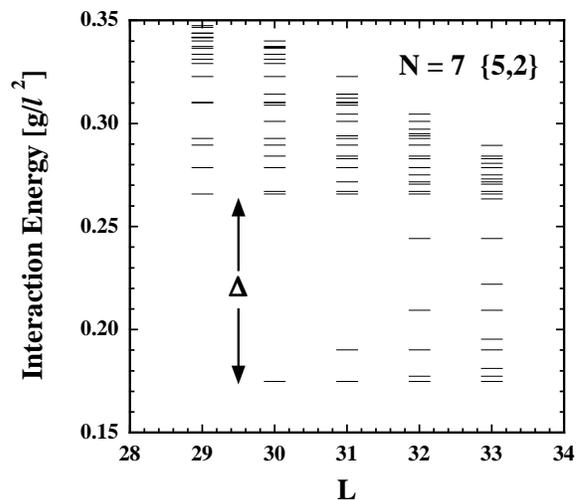}
\caption{Spectrum of the state $L = 30$ with $N = 7$. The gap $\Delta = 0.091\: g/\ell^2$.  Below
the gap, we observe two phonon branches along with their multiphonon excitations. \label{fig1}}
\end{center}
\end{figure}

{\bf Principal Jain sequence}: The incompressibility of the Laughlin droplet can be regarded
as a consequence of a statistical transmutation: at  large $N_v$,  a boson binds one vortex and 
becomes a composite object  (called composite 
fermion, CF)~\cite{WC99} that behaves as a \emph{spinless} fermion. 
In a Laughlin droplet, the CF's fill up  the $N$ lowest angular momentum orbitals in 
the  LLL. Compressing  the droplet and therefore decreasing $L$ requires promoting  CF's 
to higher LL's and costs a finite amount of energy. Thus, states with lower 
angular momentum will contain  CF's  in  higher CF LL's, and this leads to the ansatz:
\begin{equation}
\Phi(z_{1},\ldots,z_N) = {\cal P} \big[\prod_{i<j} (z_{i} - z_{j} ) \: \Phi_{\rm CF}^{[\{ N_i\} ]} ({\bf r}_1,\ldots,  {\bf r}_N)\big].
\label{CF}
\end{equation}
where ${\cal P}$ projects onto the LLL~\cite{WC99}, and $\Phi_{CF}^{[\{N_i\}]}$
is a Slater determinant with $\{N_i\}_{i=1}^p$ CF's filling the lowest angular momentum orbitals of  $p \ge 1$ 
CF  LL's and $L = L_0 = N(N-1)/2 + 
\sum_{i=0}^{p-1} N_i (N_i - (2i+1))/2$ ($N_0 = N$ and $N_i  = 0$ for $i > 0$ in the Laughlin state).  
In what follows, we focus  on the CF states $\{4,2\}$ 
and  $\{5, 2\}$, which have been shown~\cite{WC99} to have good overlap with the
exact states at $L_0 = 20$ ($N = 6$) and  $L_0 = 30$ ($N = 7$), respectively.
For a relatively large range of $L-L_0 > 0$  ($3$ for $\{4,2\}$ and $4$ for $\{5,2\}$)
the lowest energy state is a centre-of-mass excitation of the state at $L=L_0$, and the 
interaction energy is unchanged. 
Edge excitations  are those states with energy lower than the bulk gap ($\Delta$, cf. Fig.~\ref{fig1})~\cite{ZPM96}.  According 
to~\cite{W92,C03}, these states exhibit two branches of edge 
phonons described by:
\begin{equation}
H_{\rm edge} = E_0 + \sum_{m>0} \hbar \left[\omega^{(b)}_m \: b^{\dag}_m b_m 
+ \omega^{(d)}_m \: d^{\dag}_m d_m \right],~\label{ham2}
\end{equation}
and $L = L_0 + \sum_{m > 0} m \left[ b^{\dag}_m b_m + d^{\dag}_m d_m \right]$,
with $[d_m, d^{\dag}_n] =  [b_m, b^{\dag}_n] = \delta_{nm}$, commuting otherwise.
Note that  the edge excitations are not degenerate 
in energy.  However,  one can still compute the NoS  from 
$Z(q) = {\rm Tr} \: q^{L-L_0}  = \prod_{m > 0} (1 - q^m)^{-2}$. 
In  table~\ref{table1} we  compare the numerical results  
to the theoretical predictions for the NoS, finding perfect agreement   for $m= 1,2$.
For higher $m$  the results are affected by finite-size effects.
However, in the state $\{5,2\}$ the NoS for $m=3$  is  quite
close to its  $N\to \infty$  value, and therefore we concentrate on this state
for further analysis. We next  try to reproduce
the energies of the sixteen edge states of $\{5,2\}$  from $m = 1$ to $3$
using $\omega^{(b)}_{m=2}$ 
and $\omega^{(d)}_{m=1,2}$ as the only fitting parameters
($\omega^{(b)}_1$ has only a kinetic energy contribution by Kohn's theorem~\cite{C03}). 
The interaction part of $\hbar \omega^{(d)}_1$ can
be extracted from the $m = 1$ data: $\hbar \omega^{(d)}_1 = 0.015 \, g/\ell^2$.
For $m=2$, there are the five following states: 
$2^{-1/2} \: (b^{\dag}_1)^2|0\rangle$, $b^{\dag}_2 | 0\rangle$, $d^{\dag}_1 b^{\dag}_1 | 0\rangle$,
$2^{-1/2} \: (d^{\dag}_1)^2|0\rangle$, $d^{\dag}_2|0\rangle$, whose energies can be obtained from Eq.~(\ref{ham2}). For $m =3$ the  states and energies can be written down in a similar fashion. The comparison with numerics for $m = 2,3$  is given in table~\ref{table2}.  Thus, these states are good representatives of the topological order of the filling $\nu = \frac{2}{3}$ from the principal Jain sequence for bosons~\cite{W95,C03}.
\begin{table}
\caption{Interaction energies (in units of $g/\ell^2$) of the edge excitations of the $\{5,2\}$ state ($N = 7, \, L = 30$,
see Fig.~\ref{fig1}).  The predictions for multiphonon states are given in brackets.
Deviations are due to  non-linear terms not included in Eq.~(\ref{ham2}). \label{table2}}
\begin{ruledtabular}
\begin{tabular}{c|c|c||c|c}
state $\#$ &  $m=2$ (th.)  & $m=3$ (th.) & state $\#$  & $m=3$ (th.) \\ \hline  
1  & 0 (0) & 0 (0) & 6 &  0.034   (0.030) \\ 
2 & 0.002 ($\hbar \omega^{(b)}_2$) & 0.002 (0.002) & 7 & 0.047 (0.045) \\ 
3 & 0.015 (0.015) & 0.006 ($\hbar \omega^{(b)}_3$)  & 8 & 0.069 (0.069) \\
4 &  0.034 (0.030)  & 0.015 (0.015) & 9 & 0.088 (0.084) \\ 
5 &  0.069 ($\hbar \omega^{(d)}_{2}$) & 0.020 (0.017) & 10 &  
($\hbar \omega^{(d)}_3 \ge \Delta$)
\end{tabular}
\end{ruledtabular}
\end{table}

  Finally, we also find evidence (to be reported elsewhere~\cite{unpub}) for the
edge structures (three phonon branches)  corresponding to  $\nu = \frac{3}{4}$. However, 
the NoS for $m > 1$ is  affected by finite-size effects for $N= 6,\, 7$.

  {\bf Moore-Read (MR) or Pfaffian state:}   Ground states of CF's with 
 lower angular momentum   are obtained by placing 
 CF's in higher effective LL's. Eventually, when
the number of occupied levels $p \to +\infty$, the CF's would not feel any effective 
Coriolis force and the resulting state should be compressible. However, in such 
a state the CF's  can pair and condense into a BCS 
state, which would render the state incompressible again~\cite{RG00}. Since CF's 
are spinless, the pairing takes place in  $p$-wave  
and the  BCS wavefunction is a Pfaffian~\cite{RG00,MR91}, which must be multiplied
by $\prod_{i<j}(z_i-z_j)$ to yield a bosonic wavefunction:
\begin{equation}
\Phi_{\rm MR} = \prod_{i<j}(z_i-z_j) \, {\cal A} \left[\frac{1}{z_1-z_2}
 \cdots \frac{1}{z_{N-1} - z_{N}}\right],\label{MR}
\end{equation}
where $\cal A$ stands for anti-symmetrization (see \emph{e.g.} Refs.~\cite{MiR96,WG00}) of the 
bracketed  product, which is the Moore-Read (MR) wavefunction\cite{MR91}.
Soon after the MR state was introduced, it was pointed out~\cite{GWW91} that it is a zero interaction-energy  eigenstate  of a three-body potential, $U_3 = g_3 \sum_{i<j<k} \delta({\bf r}_i-{\bf r}_j)
\delta({\bf r}_i-{\bf r}_k)$. Thus,  it is convenient to work with a modified Hamiltonian:
 $H^{\prime}_{\rm LLL} = T + \lambda U_2 + (1-\lambda) U_3$, so that for $\lambda = 0$
the ground state at angular momentum $L^{\rm MR}_0 = N(N-2)/2$ 
(even $N$) or $L^{\rm MR}_0 = (N-1)^2/2$ (odd $N$) is 
\emph{exactly} the MR state. Indeed, this modification is not entirely artificial 
as one of us has recently shown that MR can become exact near a Feshbach resonance~\cite{C04}.

  For the  \emph{exact} MR state (\emph{i.e.} $\lambda = 0$) 
besides the polynomials $s_m$ introduced above, Wen~\cite{W95}, and 
Milovanovic and Read~\cite{MiR96} found a 
branch of fermionic edge excitations which are generated 
by replacing the Pfaffian in 
$\Phi_{\rm MR}$ by ${\cal A} \left[z^{n_1}_1\cdots z^{n_F}_F  
(z_{F+1} - z_{F+2})^{-1}\cdots (z_{N-1} - z_N)^{-1} \right]$, 
where $n_{1}, \ldots, n_F$ are non-negative integers. Thus the 
angular momentum  is increased by $L - L^{\rm MR}_0 = \sum_{k=1}^{F} 
\left(n_k + \frac{1}{2}\right)$. This spectrum, together with the phonon branch related to $s_m$
is described~\cite{W95} by $L = L^{\rm MR}_0 + \sum_{m > 0} \left[ m \:
b^{\dag}_m b_{m} + (m - \frac{1}{2})\: c^{\dag}_{m-\frac{1}{2}} c_{m-\frac{1}{2}} \right]$, 
where $b_m,\, b^{\dag}_m$ are the phonon operators, 
and the fermions $\{c_{m-\frac{1}{2}}, c^{\dag}_{n-\frac{1}{2}}\} = \delta_{mn}$,
anti-commuting otherwise. However, due to the paired nature of the state,   even and odd $N$ are  different. For instance,
to compute the NoS, one must define~\cite{W95} $Z^{\rm even}(q) = \frac{1}{2} 
{\rm Tr}\: (1 + (-1)^F) q^{L-L_0}$ and 
$Z^{\rm odd}(q) = \frac{1}{2} {\rm Tr}\: (1 - (-1)^F) q^{L-L_0}$, since  the parity  
$(-1)^F$, with $F = \sum_{m>0} c^{\dag}_{m-\frac{1}{2}} c_{m-\frac{1}{2}}$,
is a good quantum number~\cite{W95}. The  numerical
results are compared with the NoS for $N\to \infty$  in table~\ref{table1}.  Perfect agreement 
is found for $m \leq [N/2]$. Higher 
values of $m$ are shown to illustrate the effects of finite size; the observed differences from the $N\to \infty$ values can be also accounted for by the theory~\cite{RPC}. Furthermore, 
using the effective field theory~\cite{unpub},
we have also obtained the behavior of $n(l)$ near the edge. For $N$ even
the predicted ratios of $n(l)/n(l_{\rm max})$, where $\l_{\rm max} = N-2$, are
$1:2:3:\ldots$ For $N = 8$, we numerically find $1.0:2.2:3.4$. However, for $N$ odd
the ratios of $n(l \le l_{\rm max})/n(l_{\rm max})$ ($l_{\rm max} = N-1$) 
behave differently~\cite{unpub}:
$1:1:2:3$. For $N = 9$ we numerically find $1.0:1.0:2.5:4.1$, and the scaling  
observed from smaller systems shows  a trend of convergence to  the predicted ratios.
Although $n(l)$ had been analyzed in~\cite{W95} for a fermion MR state,
the different behavior of $n(l)$ for odd $N$ had not been described.  
\begin{figure}[t]
\begin{center}
\includegraphics[width=\columnwidth]{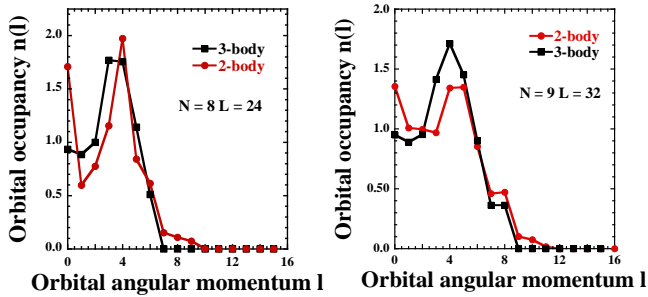}
\caption{Orbital occupancy for the Moore-Read 
(or Pfaffian) state ($3$-body) and the eigenstate of 
the $2$-body interaction identified as the Pfaffian. \label{fig2}}
\end{center}
\end{figure}

 As soon as the two-body interaction is turned on, \emph{i.e.} already
for small $\lambda > 0$,  we  observe that the \emph{plateau} at $L > L^{\rm MR}_0$ 
is lost  for small $L-L_0^{\rm MR}$.  For  pure two-body interaction ($\lambda = 1$),  
the states at $L^{\rm MR}_0 + 1$  for $N = 5, 9, 11, 13$
and  at $L^{\rm MR}_0 + 2$  for $N = 6, 7, 8, 10, 12$ have lower  
interaction energy than the ground state at $L^{MR}_0$. 
Thus, $U_2$ strongly perturbs the plateau of the exact MR state by favoring states
with angular momenta of compact CF states.
If, regardless of this fact, one counts the number of states at $L > L^{\rm MR}_{0}$
with interaction energies less than the first excited state at $L^{MR}_0$, the
results do not seem to converge and disagree with the NoS expected for the MR state 
(see table~\ref{table1}  for the observed NoS at $\lambda = 1$ for  $N = 12, 13$).We also
observe a rapid  deterioration of  the overlap of the state at $L^{\rm MR}_0$  with the exact MR state: from 
$0.91$  ($N=5$) and $0.90$ ($N=6$) to $0.68$ ($N=8$) and $0.74$ ($N=9$).
These discrepancies might be due to a very slow convergence as $N$ grows
towards a well-defined bulk MR state.  However, we note  that for very large $N$, local
density arguments~\cite{C05} show that the edge of the MR state will reconstruct. 
The anomalies  in  the edge of the small systems observed here 
are striking in view of the good behavior exhibited by the Jain states,  
which are also approximate wavefunctions.

{\bf Experimental consequences:} It is possible to excite the surface modes by
inducing a small time-dependent  deformation of the harmonic trap. 
Within linear response,   the energy injected by an $m$-polar deformation  
is  proportional to the oscillator 
strength $f_m = \sum_{\alpha} | \langle L_0+m, \alpha |O_m | L_0 \rangle |^2$,
 ($O_m = \sum^N_{i=1} z^m_i$), where $\alpha$ runs over all  edge excitations.
The theory predicts~\cite{C03} $f_m = m \nu R^{2m} $, where
$R = \sqrt{N/\nu}$ is  the droplet radius.  For  the dipole, $f_1 = N$, as required by  
Kohn's theorem~\cite{C03}, and confirmed by our numerics. For the quadrupole ($m = 2$) 
$f_2 = 2 N^2/\nu$. We have numerically tested the accuracy of this formula for the $m=2$ modes of the 
states in table~\ref{table1} (except the MR state at $\lambda = 1$), finding that already for $N = 5,6$  the deviations are no larger than $7\%$.  
This result suggests an experimental way of estimating the filling fraction,
provided $f_2$ and $N$ (or $f_1$) can be measured: $\nu^{-1}_{\rm exp} = f_2/2 N^2$.
More details will be provided elsewhere~\cite{unpub}.

  MAC thanks  M. Greiter for a useful conversation  and the department 
ECM of  the \emph{Universitat de Barcelona} for  hospitality. We are
also grateful to N. Read for helpful  correspondence. 
Financial support from a \emph{Gipuzkoa} 
fellowship (MAC) of the 
\emph{Gipuzkoako Foru Aldundia}  and  grants (NB) no. BFM2002-01868
from DGESIC (Spanish Goverment) and  2001SGR00064 (Generalitat de Catalunya)
and (NRC) GR/S61263/01 (EPSRC) is  gratefully acknowledged.


\begin{thebibliography}{30}
%
\bibitem{WG00}
N. Wilkin and J.~M.~F. Gunn, Phys.~Rev.~Lett. {\bf 84}, 6  (2000).
%
\bibitem{CWG01}
N.~R. Cooper, N.~K. Wilkin, and J.~M.~F. Gunn, Phys.~Rev.~Lett.~{\bf 87} 120405 (2001).
%
\bibitem{B04}
See G. Baym,  cond-mat/0408401, for an up-to-date review
of the recent progress in the field. 
%
\bibitem{SC04}
V. Schweikhard \emph{et al.}, Phys. Rev. Lett. {\bf 92}, 040404 (2004).
%
\bibitem{Bt04}
V. Bretin \emph{et al.}, Phys.~Rev.~Lett.~{\bf 92}, 050403 (2004).
%
\bibitem{W95}
X.-G. Wen, Adv. Phys.  {\bf 44}, 405 (1995); Phys.~Rev.~Lett.~{\bf 70}, 355
(1992).
%
\bibitem{W92}
X.-G. Wen, Int.~J. of Mod.~Phys. {\bf 6}, 1711 (1992).
%
\bibitem{C03}
M.~A. Cazalilla, Phys.~Rev.~A {\bf 67} 063613 (2003).
%
\bibitem{W02}
X.-G. Wen,  Physics Letters A {\bf 300}, 175 (2002).
%
\bibitem{Ho01}
N. K. Wilkin, J. M. F. Gunn, and R. A. Smith, Phys.~Rev.~Lett.~{\bf 80} 2265 (1998); T.~L. Ho, Phys.~Rev.~Lett.~{\bf 87} 060403 (2001).
%
\bibitem{WC99}
 N.~R. Cooper and N.~K.~ Wilkin, Phys.~Rev.~B {\bf 60} R16279 (1999).
%
\bibitem{RJ03}
N. Regnault and Th. Jolicoeur, Phys.~Rev.~Lett. {\bf 91}, 030402 (2003).
%
\bibitem{ZPM96}
U. Z\"ulicke, J.~J. Palacios, and A.~H. MacDonald, Phys.~Rev.~B~{\bf 67},045303 (2001).
%
\bibitem{BOL04}
To increase the gaps,  rotating dipolar gases has been recently proposed. See 
M.~A. Baranov, K. Osterloh, and M. Lewenstein,  cond-mat/0404329.
%
\bibitem{RG00}
N.~Read and D. Green, Phys.~Rev.~ B {\bf 61}, 10267 (2000).
%
\bibitem{MR91}
G. Moore and N. Read, Nucl.~ Phys. B{\bf 360}, 362 (1991).
%
\bibitem{GWW91}
M. Greiter, X.~G. Wen, and F. Wilczek, Phys.~Rev.~Lett.~{\bf 66}, 3205 (1991).
%
\bibitem{MiR96}
M. Milovanovic and N. Read, Phys.~Rev.~B {\bf 53}, 13559 (1996).
%
\bibitem{C04}
N.~R. Cooper,  Phys.~Rev.~Lett. {\bf 92}, 220405 (2004).
%
\bibitem{RPC}
N. Read, \emph{private communication}; N. Read and E.~H. Rezayi, Phys.~Rev.~B {\bf 54},
16864 (1996).
%
\bibitem{C05}
N.~R. Cooper \emph{et al.}, cond-mat/0409146.
%
\bibitem{unpub}
M.~A. Cazalilla, N. Barber\'an, and N.~R. Cooper, in preparation.
%
\end{thebibliography}
\end{document}